\documentclass{aa}  

\usepackage{graphicx}
\usepackage{txfonts}
\usepackage{color}
\usepackage{threeparttable}
\usepackage{natbib}
\bibpunct{(}{)}{;}{a}{}{,}

\newcommand{\oergcm}[1]{$10^{#1}$ erg cm$^{-2}$ s$^{-1}$}
\newcommand{\ergs}[1]{$\times 10^{#1}$ erg s$^{-1}$}
\newcommand{\erg}[1]{$10^{#1}$ erg s$^{-1}$}
\newcommand{\oergs}[1]{$10^{#1}$ erg s$^{-1}$}

\newcommand{\ohcm}[1]{$10^{#1}$ cm$^{-2}$}

\newcommand{\cts}{cts s$^{-1}$\xspace}

\newcommand{\ltsima}{$\buildrel < \over \sim$}
\newcommand{\lsim}{\lower.5ex\hbox{\ltsima}}
\newcommand{\gtsima}{$\buildrel > \over \sim$}
\newcommand{\gsim}{\lower.5ex\hbox{\gtsima}}
\newcommand{\msun}{M$_{\odot}$}

\newcommand{\xmm}{{\it XMM-Newton}\xspace}

\newcommand{\src}{\hbox{XMMU\,J053108.3$-$690923}\xspace}

\usepackage[normalem]{ulem}

\begin{document} 

\title{Fast flaring observed from XMMU\,J053108.3$-$690923 by eROSITA: a supergiant fast X-ray transient in the Large Magellanic Cloud}

\author{C. Maitra\inst{1} \and
        F. Haberl\inst{1} \and
        G. Vasilopoulos\inst{2} \and
        L. Ducci\inst{3} \and
        K. Dennerl\inst{1} \and
        S. Carpano\inst{1}
       } 

\titlerunning{SXFT in LMC}
\authorrunning{Maitra et al.}

\institute{
  Max-Planck-Institut f{\"u}r extraterrestrische Physik, Gie{\ss}enbachstra{\ss}e 1, 85748 Garching, Germany, \email{cmaitra@mpe.mpg.de}
  \and
  Department of Astronomy, Yale University, PO Box 208101, New Haven, CT 06520-8101, USA
  \and
  Institut f{\"u}r Astronomie und Astrophysik, Kepler Center for Astro and Particle Physics, Eberhard Karls Universit{\"a}t, Sand 1, 72076 T{\"u}bingen, Germany
  }

\date{Received ... / Accepted ...}

\abstract{
   {Context. Supergiant fast X-ray transients (SFXTs) are a peculiar class of supergiant high-mass X-ray binary (HMXB) systems characterised by extreme variability in the X-ray domain. In current models, this is mainly attributed to the clumpy nature of the stellar wind coupled with gating mechanisms involving the spin and magnetic field of the neutron star.}
   
   {Aims. We studied the X-ray properties of the supergiant HMXB \src in the Large Magellanic Cloud to understand its nature.}
   
   {Methods. We performed a detailed temporal and spectral analysis of the eROSITA and \xmm data of \src.}
  
   {Results. We confirm the putative pulsations previously reported for the source with high confidence, certifying its nature as a neutron star in orbit with a supergiant companion. We identify the extremely variable nature of the source in the form of flares seen in the eROSITA light curves. The source flux exhibits a total dynamic range of more than three orders of magnitude, which confirms its nature as an SFXT, and is the first such direct evidence from a HMXB outside our Galaxy exhibiting a very high dynamic range in luminosity as well as a fast flaring behaviour. We detect changes in the hardness ratio during the flaring intervals where the hardness ratio reaches its minimum during the peak of the flare and increases steeply shortly afterwards. This is also supported by the results of the spectral analysis carried out at the peak and off-flare intervals. This scenario is consistent with the presence of dense structures in the supergiant wind of \src where the clumpy medium becomes photoionised at the peak of the flare leading to a drop in the photo-electric absorption. Further, we provide an estimate of the clumpiness of the medium and the magnetic field of the neutron star assuming a spin equilibrium condition.}
   {}}


\maketitle   


\section{Introduction}
\label{sec:intro}
Supergiant fast X-ray transients (SFXTs) are an elusive class of sources amongst the `classic' supergiant high-mass X-ray binaries hosting a neutron star (NS)  that is accreting from the wind of an O-B supergiant companion 
\cite[see][and references therein]{ROMANO2015126,Bird_2016,2017A&A...608A.128B,2017SSRv..212...59M}. 
Unlike the classical systems that exhibit luminosity variations in the range of 10--50, SFXTs are characterised by a larger dynamic range in X-ray luminosity ($10^{2}-10^{5}$) and exhibit fast flaring activity on timescales of a few hours.
These sources spend most of their lifetime in quiescence at luminosities of $10^{32}-10^{33}$\,erg\,s$^{-1}$ and undergo sporadic flares/outbursts reaching luminosities of $10^{36}-10^{37}$\,erg\,s$^{-1}$ \citep{2011A&A...531A.130B,ROMANO2015126,refId0}.
This remarkable variability was originally attributed to accretion from the extremely clumpy stellar wind of the companion in large eccentric orbits around the compact object \citep{2005A&A...441L...1I,2010ASPC..422...57N}.  Similar systems were subsequently discovered, but with shorter orbital periods and/or less-dense wind. This has led to the proposition of several alternative scenarios. One proposes that SFXTs host a slowly spinning highly magnetised neutron star with the large luminosity swings explained by transitions across the magnetic and/or centrifugal barrier \citep{Bozzo_2008}. Alternately, \cite{Grebenev2007149} propose an explanation of the SFXT behaviour by transitions across the centrifugal barrier by fast-rotating neutron stars (spin period shorter than $\sim$3\,s), assuming a mass-loss rate of 10$^{-5}$ solar \msun \,yr$^{-1}$, relative velocity of 1000\,km s$^{-1}$, circular orbit with a period of 5 days, and a typical NS surface magnetic field of $10^{12}$\,G. More recently, the large dynamic luminosity range of SFXTs  has been explained by the system residing in the subsonic accretion regime most of the time, with the flares being triggered by magnetic reconnections between the NS and the magnetised wind of the supergiant \citep{10.1111/j.1365-2966.2011.20026.x,10.1093/mnras/stu1027}. 
The Magellanic Clouds are a treasure trove of high-mass X-ray binaries (HMXBs), given the high formation efficiency, relatively short distance, and low foreground absorption, all conducive to performing detailed studies. The Small Magellanic Cloud in particular hosts the largest sample of Be/X-ray binaries known in any galaxy \citep{2016A&A...586A..81H}. On the other hand, the population of HMXBs in the Large Magellanic Cloud (LMC) is relatively unexplored given its large angular size on the sky and a previously incomplete coverage in X-rays at energies above $\sim$2 keV. A total of $\sim$60 HMXBs have been found in the LMC, of which $\sim$90\% are Be/X-ray binaries, with the rest being supergiant systems \citep{2020ATel13609....1H,2020ATel13610....1M,2019MNRAS.490.5494M,2018MNRAS.475.3253V,2016MNRAS.459..528A}, while one of these systems has been suggested to be a probable SFXT based on its fast flaring properties \citep[\src, ][]{2018MNRAS.475..220V}. 

XMMU\,J053108.3$-$690923 is a high-mass X-ray binary pulsar recently identified in the LMC based on its optical and X-ray properties. The optical companion was identified as a B0 IIIe star by \cite{2018MNRAS.475.3253V} based on optical spectroscopy. \cite{2018MNRAS.475..220V} further classified the system as a bright giant or supergiant B0 IIe-Ib based on a more robust classification using a combination of spectroscopic and hybrid spectrophotometric criteria. Tentative pulsations of $\sim$2013\,s were detected from an \xmm observation performed on 2012-10-10 (Obsid 0690744601) when the source was found,  serendipitously, to be in the field of view \citep{2018MNRAS.475..220V}. Here we present a detailed X-ray timing and spectral analysis of the source using eROSITA observations and  a recent dedicated \xmm observation. In this work we identify \src 
as a fast-flaring SFXT, the first  from outside our Galaxy for which there is such direct evidence. Section~\ref{sec:observations} presents our observational data, Sect.~\ref{sec:results} our results, and Sect.~\ref{sec:discussion} a discussion and our conclusions.

\section{Observational data}
\label{sec:observations}
In several long eROSITA pointed observations during the commissioning and calibration phase, \src was in the field of view of selected cameras.
The source was also scanned during the first all-sky survey (eRASS:1). 
The eROSITA observations are summarised in Table~\ref{tabobs}. 
Additionally, a dedicated \xmm observation of \src was performed in 2018 (Table~\ref{tab:observations}).
\begin{table*} 
\caption{eROSITA Observations of \src} 
\begin{center}
\begin{tabular}{cccc} 
\hline\hline
Telescope module & Time (UTC)      & Net exposure (ks) & Vignetting (0.6-2.3\,keV)\\
\hline
 \multicolumn{3}{c}{SN 1987A commissioning (Obs.\,1, Obsid 700016)} \\
 TM3 &  2019-09-15 09:23:06--2019-09-16 17:00:10  & 103 & 0.52\\ 
 TM4 &  2019-09-15 19:46:07--2019-09-16 17:00:11  &  76 & 0.50\\
\hline
 \multicolumn{3}{c}{SN 1987A first-light (Obs.\,2, Obsid 700161)} \\
 TM1 &  2019-10-18 16:54:34--2019-10-19 15:07:54 & 78.9 & 0.45\\ 
 TM2 &  2019-10-18 16:54:34--2019-10-19 15:07:54 & 79.7 & 0.54\\
 TM3 &  2019-10-18 16:54:34--2019-10-19 15:07:54 & 79.7 & 0.57\\
 TM4 &  2019-10-18 16:54:34--2019-10-19 15:07:54  & 79.7 & 0.54\\
 TM6 &  2019-10-18 16:54:34--2019-10-19 15:07:54 & 79.7 & 0.53\\
\hline
 \multicolumn{3}{c}{N132D  (Obs.\,3, Obsid 700183)}   \\
 TM1 & 2019-11-25 19:13:51--2019-11-26 06:20:31 & 40.0 & 0.44\\ 
 TM2 & 2019-11-25 19:13:51--2019-11-26 06:20:31  & 27.0 & 0.45\\
 TM3 & 2019-11-25 19:13:51--2019-11-26 06:20:31  & 27.0 & 0.46\\
 TM4 & 2019-11-25 19:13:51--2019-11-26 06:20:31  & 8.0 & 0.46\\
 TM6 & 2019-11-25 19:13:51--2019-11-26 06:20:31 & 40.0 & 0.45\\
\hline 
\multicolumn{3}{c}{eRASS:1  (Obs.\,4)}   \\
 TM1 & 2020-04-26 08:57:34--2020-05-07 16:57:41 & 2.3 & 0.47\\ 
 TM2 & 2020-04-26 08:57:34--2020-05-07 16:57:41  & 2.3 & 0.48\\
 TM3 & 2020-04-26 08:57:34--2020-05-07 16:57:41  & 2.1 & 0.48\\
 TM4 & 2020-04-26 08:57:34--2020-05-07 16:57:41  & 2.2 & 0.47\\
 TM6 & 2020-04-26 08:57:34--2020-05-07 16:57:41 & 2.3& 0.48\\

\label{tabobs} 
\end{tabular} 
\end{center}
\end{table*}



\begin{table*}
\centering
\caption[]{\xmm observation of \src.}
\begin{tabular}{lccccrrr}
\hline\hline\noalign{\smallskip}
\multicolumn{1}{c}{Observation} &
\multicolumn{1}{c}{Date} &
\multicolumn{1}{c}{Exposure time} &
\multicolumn{1}{c}{Off-axis} &
\multicolumn{1}{c}{R.A.} &
\multicolumn{1}{c}{Dec.} &
\multicolumn{1}{c}{Err} \\
\multicolumn{1}{c}{ID} &
\multicolumn{1}{c}{} &
\multicolumn{1}{c}{pn, MOS1, MOS2} &
\multicolumn{1}{c}{angle} &
\multicolumn{1}{c}{(J2000)} &
\multicolumn{1}{c}{(J2000)} &
\multicolumn{1}{c}{} \\
\multicolumn{1}{c}{} &
\multicolumn{1}{c}{} &
\multicolumn{1}{c}{(s)} &
\multicolumn{1}{c}{(\arcmin)} &
\multicolumn{1}{c}{(h m s)} &
\multicolumn{1}{c}{(\degr\ \arcmin\ \arcsec)} &
\multicolumn{1}{c}{(\arcsec)} \\
\noalign{\smallskip}\hline\noalign{\smallskip}
 0822310101 & 2018-01-24 & 45000, 31347, 31494 & 1.0 & 5 31 08.31 & -69 09 25.6 & 0.4 \\
\noalign{\smallskip}\hline
\end{tabular}
\tablefoot{
A bore-sight correction using four QSOs was applied to the coordinates.
Net exposure times after background-flare screening for pn, MOS1, and MOS2, respectively.
The off-axis angle is given for the EPIC-pn telescope.
}
\label{tab:observations}
\end{table*}

\subsection{eROSITA}
During the commissioning phase of eROSITA (Predehl et al. 2020) SN\,1987A was observed on 2019-09-15 (hereafter referred to as Obs.\,1) with only the cameras from telescope modules 3 and 4 (TM3, TM4) observing the sky with \src in the field of view. The observation duration was $\sim$100\,ks. 
During the eROSITA first-light observation targeted on SN\,1987A on 2019-10-18 (Obs.\,2), the source was in the field of view, but was too faint for a detailed timing and spectral analysis to be performed. 
In a calibration and performance-verification (Cal/PV) observation targeted on the supernova remnant N132D (Obs.\,3) all seven TMs observed the sky  and collected data. However, for the timing and spectral studies, only the telescope modules with on-chip filter are used (TM1, TM2, TM3, TM4 and TM6), because 
no reliable energy calibration is available for the others (TM5 and TM7) due to an optical light leak.
The source was also detected during eRASS1 (Obs.\,4) for several scans. Table~\ref{tabobs} provides the observation details.

The data were reduced with a  pipeline based on the eROSITA Standard Analysis Software System \citep[eSASS,][]{Predehl2020},
which determines good time intervals, corrupted events and frames, and dead times, masks bad pixels, and applies pattern recognition and energy calibration. Finally, star-tracker and gyro data are used to assign celestial coordinates to each reconstructed X-ray photon. TM4 data for Obs.\,3 were not used for spectral analysis, because after filtering it had only an effective exposure of $\sim$8\,ks. Circular regions with a radius of 32\arcsec\ and 50\arcsec\ were used for the source and background extraction, respectively.

Source detection was performed simultaneously on all the images in the standard eROSITA energy bands of 0.2--0.6\,keV, 0.6--2.3\,keV, and 2.3--5.0\,keV. The source detection is based on a maximum likelihood point spread function (PSF) fitting algorithm that determines best-fit source parameters and detection likelihoods. Data from all the seven telescope modules were used for this purpose.
The vignetting and PSF-corrected count rates in the energy range of 0.2-5.0\,keV for Obs.\,1, 2, 3, and 4 are 0.130$\pm$0.004 \cts, 0.005$\pm$0.001 \cts, 0.180$\pm$0.005 \cts and 0.049$\pm$0.008 \cts, respectively.
The count rates mentioned in this paper using eROSITA are all normalised to 7 TMs. The best-fit determined  position from eROSITA after a preliminary astrometric correction is R.A. = 05$^{\rm h}$31$^{\rm m}$07.95$^{\rm s}$ and Dec. = $-$69\degr09\arcmin25\farcs4 (J2000) with a $1\sigma$ statistical uncertainty of 2.3\arcsec. This position is consistent within errors with that inferred from the \xmm data.

\subsection{\xmm}
After the serendipitous discovery of \src within the field of view \citep{2018MNRAS.475..220V}, the source was followed up with a dedicated \xmm pointing (obsid 0822310101, 45\,ks exposure, PI: Vasilopoulos). \xmm/EPIC \citep{2001A&A...365L..18S,2001A&A...365L..27T} data were processed using the latest \xmm data analysis software SAS, version 18.0.0\footnote{Science Analysis Software (SAS): \url{http://xmm.esac.esa.int/sas/}}. Observations were inspected for high background-flaring activity by extracting the high-energy light curves (7.0\,keV$<$E$<$15\,keV for pn and both MOS detectors) with a bin size of 100\,s. Event extraction was performed with the SAS task \texttt{evselect} using the standard filtering flags (\texttt{\#XMMEA\_EP \&\& PATTERN<=4} for EPIC-pn and \texttt{\#XMMEA\_EM \&\& PATTERN<=12} for EPIC-MOS).
Circular regions with a radius of 15\arcsec\ and 30\arcsec\ were used for the source and background extraction to maximise the number of extracted photons. Source detection was performed  simultaneously on all the images using the SAS task {\tt edetect\_chain} in the standard \xmm energy bands. The best-fit position and the error are tabulated in Table~\ref{tab:observations}.
The source was detected with a net count rate of 0.018$\pm0.001$ \cts in the energy range of 0.2--12.0\,keV (pn+M1+M2).


\section{Results}
\label{sec:results}
\subsection{X-ray timing analysis}
The eROSITA light curves for Obs.\,1, 3, and 4 were corrected for vignetting and PSF losses by the eSASS task \texttt{srctool}.  The observation Obs.\,2 was of insufficient statistical quality to perform a meaningful timing analysis. The \xmm task \texttt{epiclccorr} was used for the same purpose for the \xmm/EPIC light curves.
 To look for the periodic signal in the X-ray light curve of \src as reported in \cite{2018MNRAS.475..220V}, a Lomb-Scargle periodogram analysis in the period range of 10-3000\,s \citep{1976Ap&SS..39..447L,1982ApJ...263..835S} was performed.

\label{subsec:timing}
\subsubsection{Obs.\,1 }
\begin{figure*}
   \centering
  \includegraphics[scale=0.6]{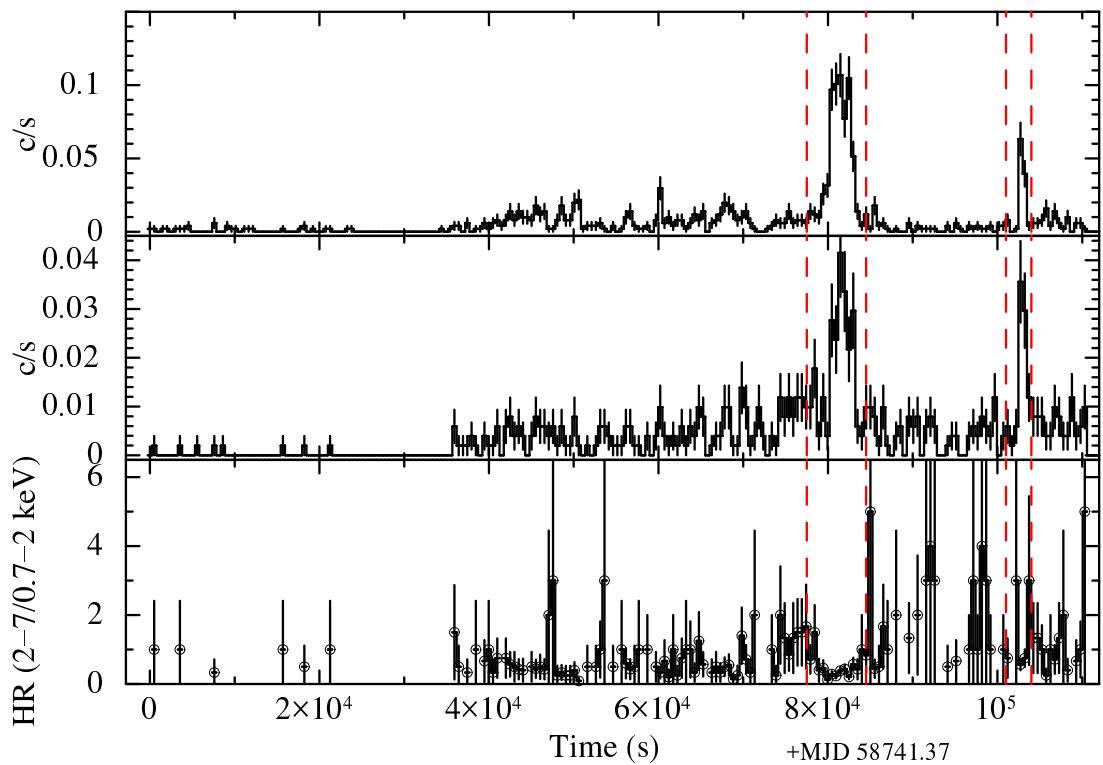}
\caption{Top: Combined eROSITA (TM3 and 4) light curve for Obs.\,1 in two energy bands 0.7--2\,keV (top) and 2.0--7.0\,keV (middle) and the corresponding hardness ratio (bottom). Both light curves are binned with 506\,s. The two flare intervals are marked with dashed vertical lines. Count rates are normalised to 7 TMs.}
 \label{figcommlc}
\end{figure*}

The light curve of \src binned at 506\,s (one-quarter of the pulse period, see below) shows strong flaring activity with the source remaining in a fainter state for most of the observation  (Fig.~\ref{figcommlc}). A large flare is seen around MJD 58742.35 lasting for $\sim$2.5\,hr, followed by another smaller flare lasting for $\sim$50\,min around MJD 58742.61. During the larger flare, the intensity increases by a factor of $30\pm8$ (0.7--2\,keV) compared to the mean count rate observed during the interval between the two flares. Comparison of the light curve in the two energy bands (0.7--2.0\,keV and 2.0--7.0\,keV) and computation of the corresponding hardness ratio  indicates that the flares are stronger in the soft X-ray band (i.e. 0.7--2.0\,keV) with the larger flare being the softest in nature. Further, the hardness ratio increased immediately after the flare in both cases, indicating a hardening of the spectrum (Fig.~\ref{figcommlc} bottom panel). This is also supported by the spectral analysis, as shown in Sect.~\ref{subsec:specr}.

\begin{figure*}
   \centering
  \includegraphics[scale=0.95]{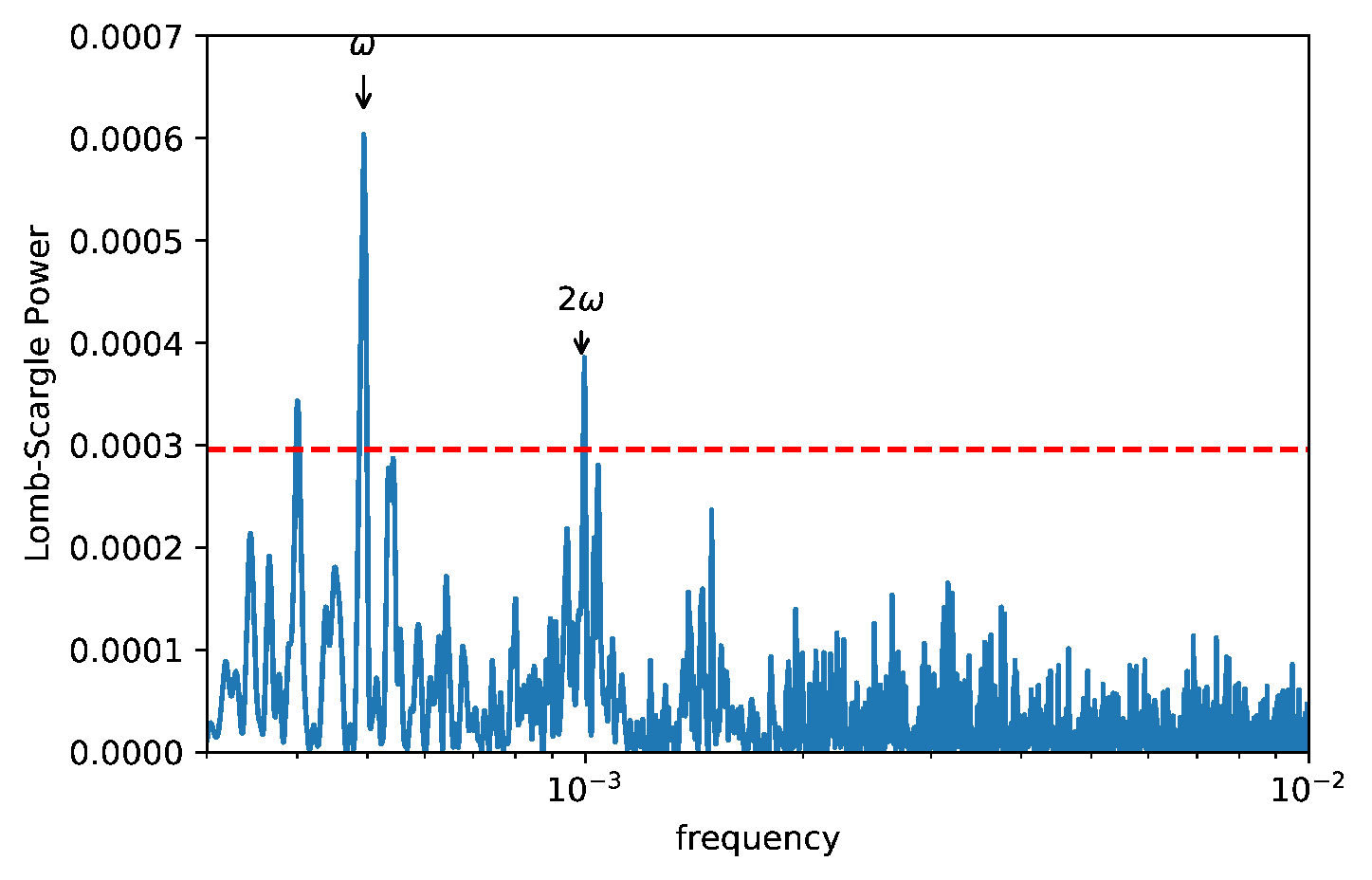} 
  \caption{Lomb-Scargle periodogram of the combined eROSITA (TM3 and TM4) light curve in the energy band of 0.2--10\,keV (Obs.\,1). The peak indicates the spin period of the neutron star of $2025\pm27$\,s. The red line marks the 90\% confidence level.}
   \label{figperobs1}
\end{figure*}

In the case of Obs.\,1, a strong periodic signal with the fundamental at $\sim$2024\,s as well as its first harmonic are visible in the periodogram, confirming the earlier reported spin period of the neutron star (Fig.~\ref{figperobs1}). The significance of the pulsations decreased with the removal of the flare intervals from the light curve. An epoch-folding technique was applied to determine the spin period more precisely  \citep{1987A&A...180..275L}. The corresponding value of the spin period and its 1$\sigma$ error is $2025\pm27$\,s. The value of the pulse period is consistent within errors with the spin period derived from the earlier \xmm observation \citep{2018MNRAS.475..220V}. Figure~\ref{figcommpf} (left panel) shows the background-subtracted and corrected light curve in the energy range of 0.2--10.0\,keV folded with the best-fit period. The profile reveals a sharp single peak with a pulsed fraction, computed as the ratio of $(I_{max}-I_{min})/(I_{max}+I_{min})$, which is equal to $\sim$40\%. Pulsations are detected with higher significance in the hard X-ray band with a pulsed fraction of $\sim$60\% in the energy range of 2.0--7.0\,keV.


\subsubsection{Obs.\,3}

\begin{figure*}
\hspace*{1,2cm}\includegraphics[scale=0.6]{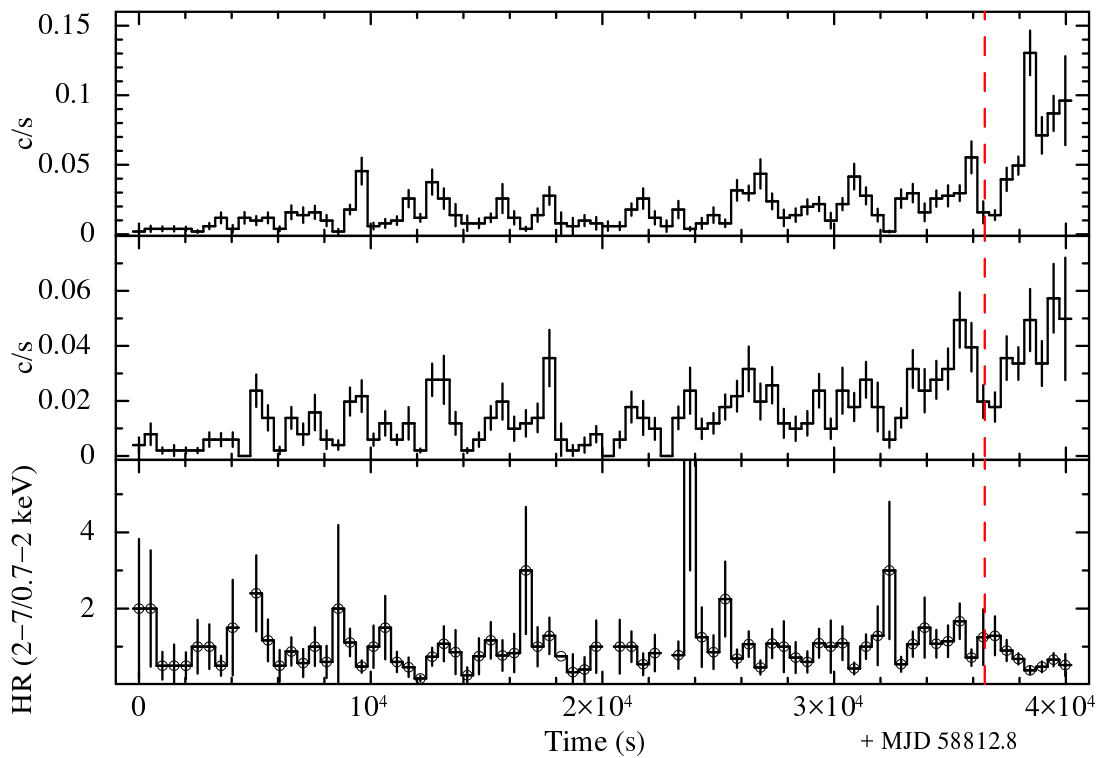}
\caption{Same as in Fig.~\ref{figcommlc} for Obs.\,3. The flare interval is marked with dashed vertical lines.}
 \label{figcomm2lc}
\end{figure*}

The light curve of \src binned at 506\,s  shows a  flare lasting for $\sim$80\,min towards the end of the observation around MJD 58813.22 where the intensity increases by a factor of $15\pm2$ compared to the mean count rate observed during the rest of the interval (Fig.~\ref{figcomm2lc}).
The light curves in two energy bands (0.7--2.0\,keV and 2.0--7.0\,keV) and the  corresponding hardness ratio  indicate that the flare is stronger in the soft X-ray band (i.e. 0.7--2.0\,keV) as was the case in Obs.\,1. 

\begin{figure*}
   \centering
  \includegraphics[scale=0.85]{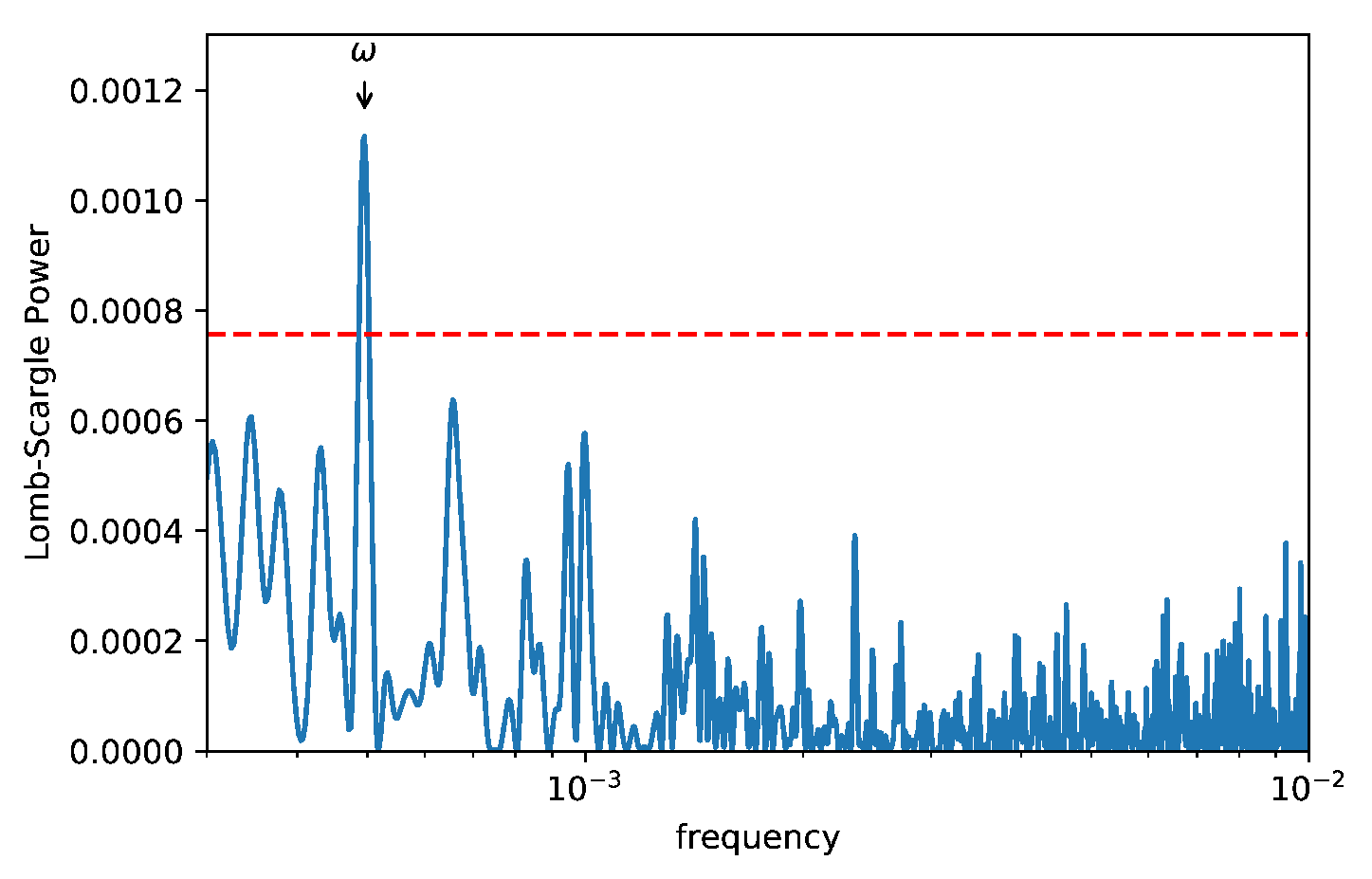}
  \caption{Lomb-Scargle periodogram of the combined eROSITA light curve in the energy band of 0.2--10\,keV (Obs.\,3) with the flare interval removed. The arrow indicates the spin period of $2018\pm29$\,s. The red dashed line marks the 90\% confidence level.}
   \label{figperobs3}
\end{figure*}

Pulsations are not significantly detected in the light curve from the entire observation. Pulsations are however detected after removing the flare interval as shown in Fig.~\ref{figperobs3}.  Using an epoch-folding technique, the best-fit spin period obtained is 2018$\pm29$\,s. The pulse profile exhibits a similar morphology to that seen in the previous eROSITA and the \xmm observations and displays a similar pulsed fraction of $\sim$40\% over the entire energy band (Fig.~\ref{figcommpf}). Again, the pulsations are detected with higher significance in the hard X-ray band (2.0--7.0\,keV) with a pulsed fraction of $\sim$70\%. No significant energy dependence of the pulse shape can be detected in the 0.2--10.0\,keV energy band. 
\subsubsection{Obs.\,4}
\begin{figure*}
   \centering
  \includegraphics[scale=0.74]{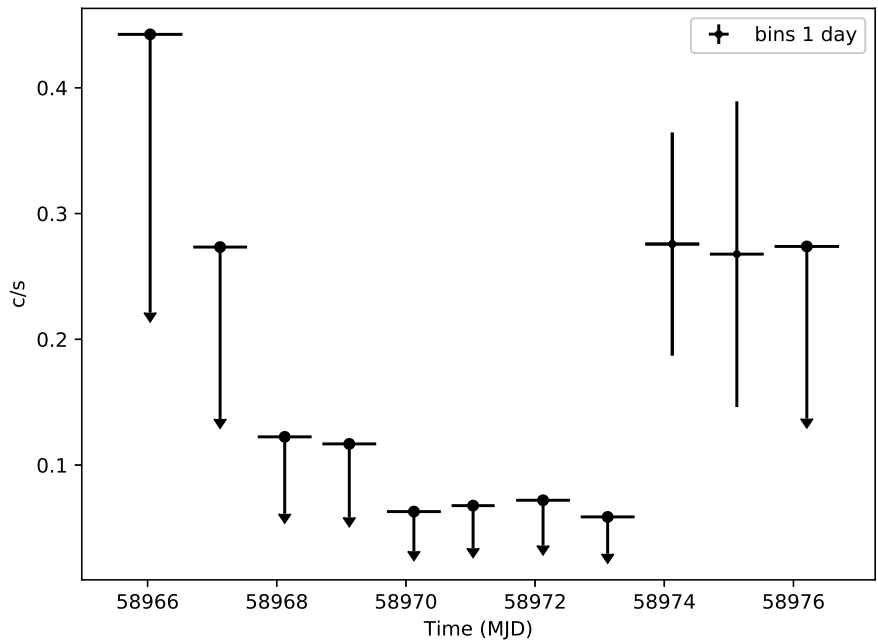}
\caption{eRASS:1 light curve of \src showing day-to-day variations of the source. Each bin includes six or seven scans of typically 30-40\,s exposure. Count rates are normalised to 7 TMs. The upper limits correspond to 2$\sigma$ detection.}
\label{figlcerass}
\end{figure*}

Due to its proximity to the South Ecliptic Pole, \src was scanned 71 times  during
eRASS1 covering the time range between 2020-04-26 08:57:34 and 2020-05-07 16:57:41. The effective exposure of the source during eRASS:1 is $\sim$2.2\,ks. Figure~\ref{figlcerass} shows the variation of the source averaged over about 1 day. As is evident in the figure, the source was faint most of the time with a count rate of $0.05$ \cts  (0.2--10\,keV), which indicates a flux that is $\sim$20 times lower than during the eROSITA Cal/PV observations. Signatures of flaring activity are however seen between MJD 58974--58976
when the count rate rose to 0.3$\pm0.1$ \cts (0.2--10\,keV) which corresponds to a luminosity of 2$\times$\oergs{36} assuming the same spectral parameters as inferred from the previous eROSITA observations, and a distance of 50\,kpc.
\subsection{\xmm}
\begin{figure*}
   \centering
 \includegraphics[scale=0.5,angle=-90]{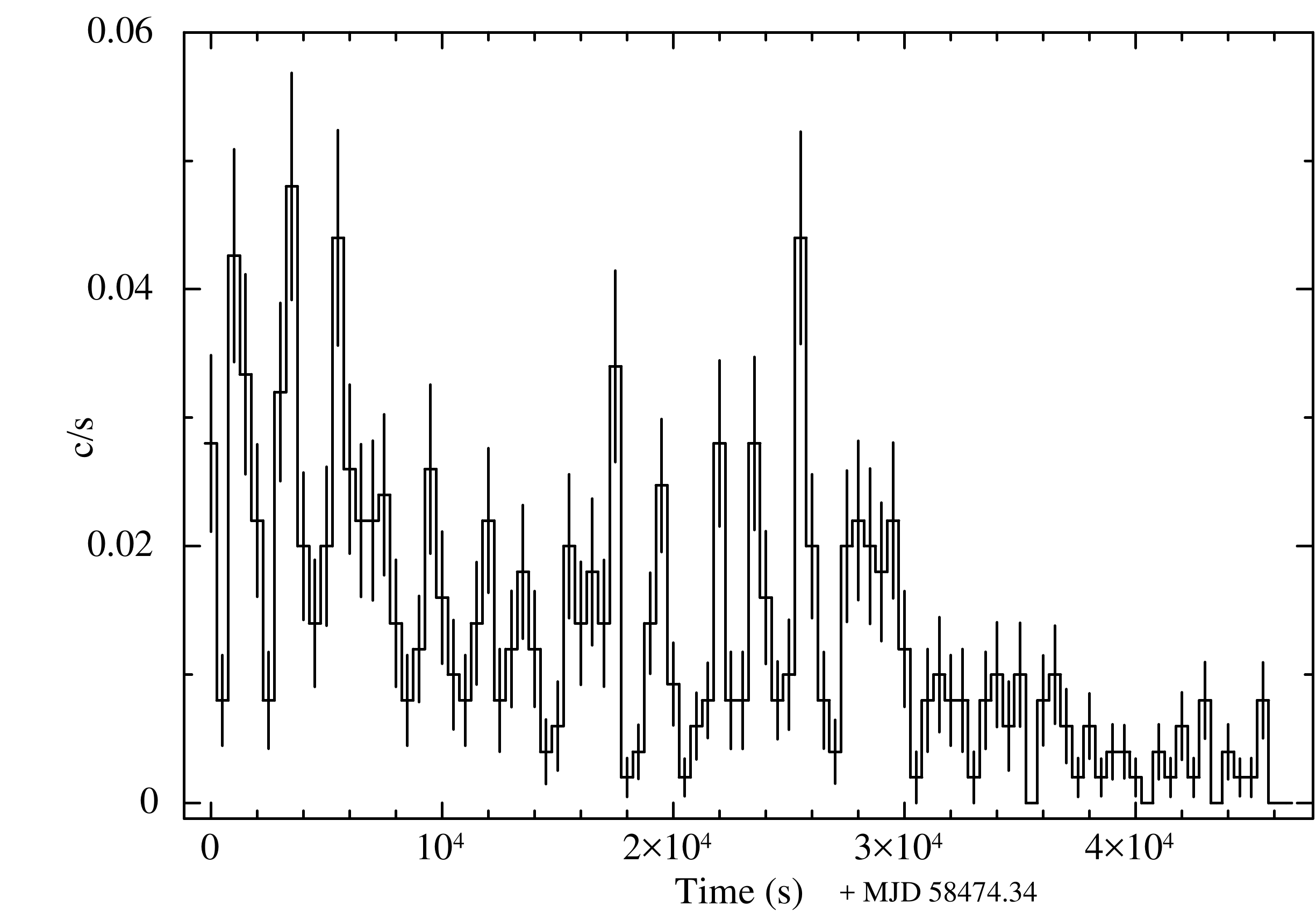} \\
\caption{Combined EPIC light curve for  \src using data from the \xmm observation 0822310101.}
   \label{figlcxmm}
\end{figure*}

\begin{figure*}
\centerline{
  \hspace{-0.6cm}
  \includegraphics[width=0.31\textwidth,angle=-90]{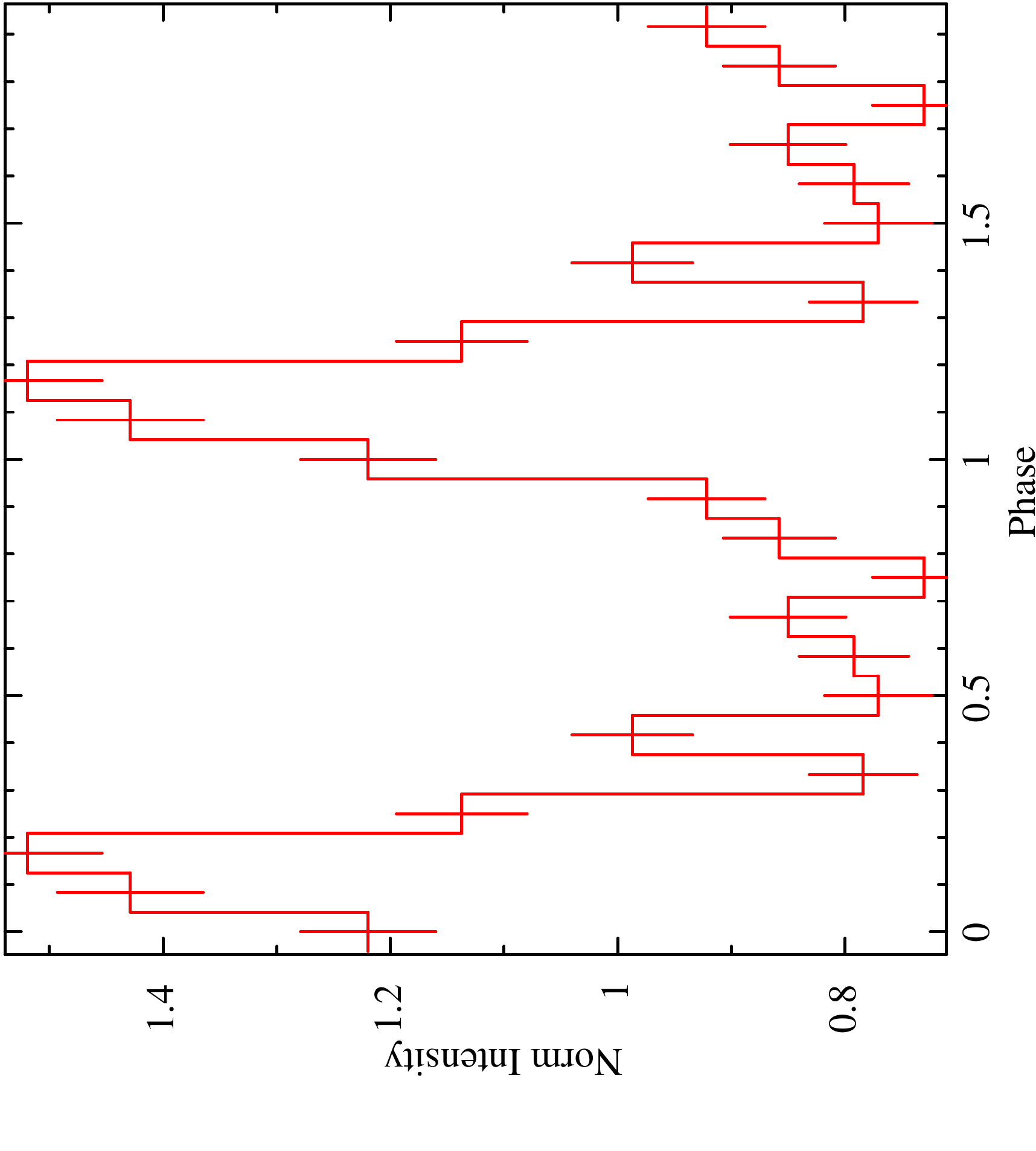} 
  \hspace{-0.3cm}
  \includegraphics[width=0.31\textwidth,angle=-90]{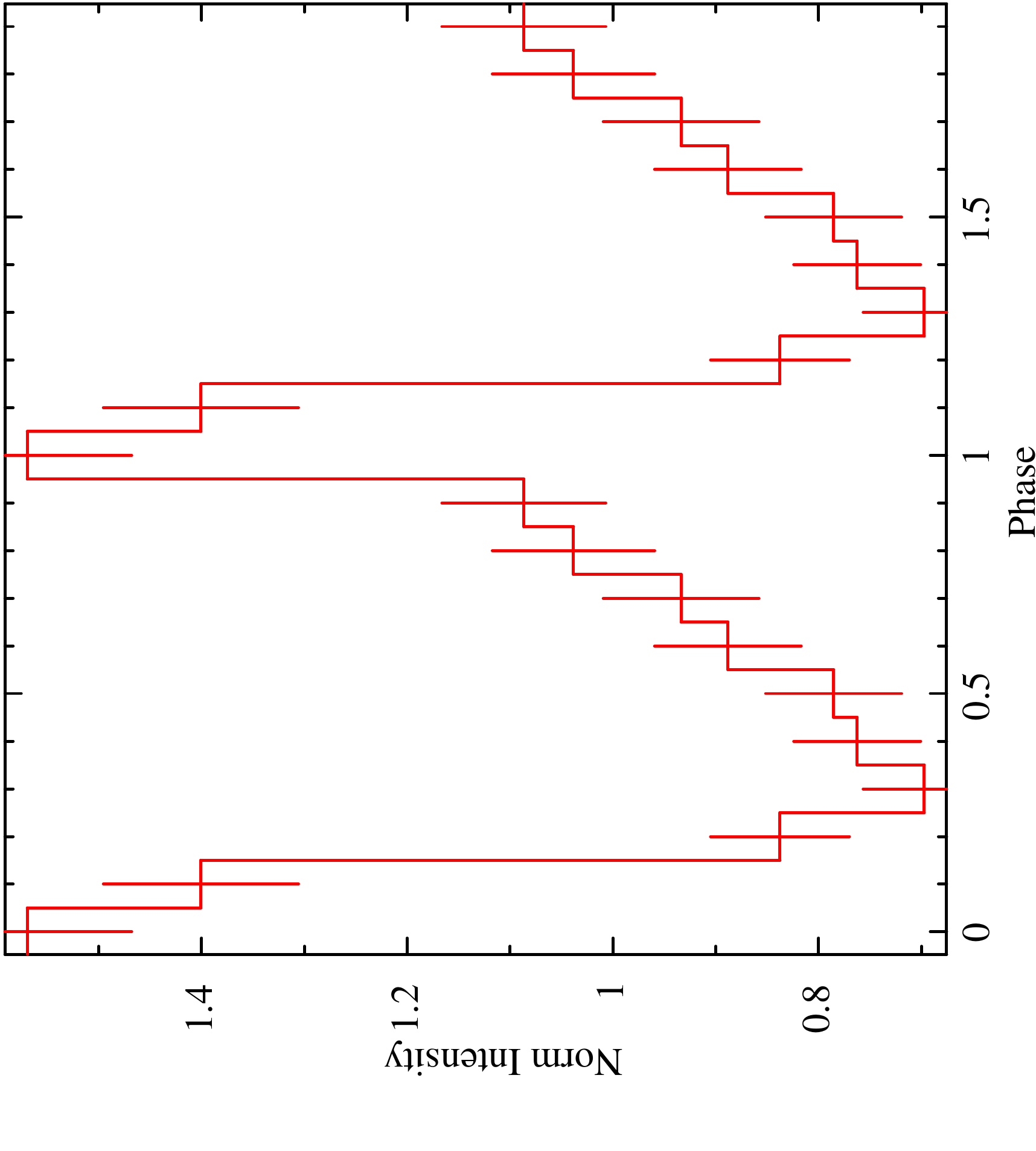}
  \hspace{-0.3cm}
  \includegraphics[width=0.31\textwidth,angle=-90]{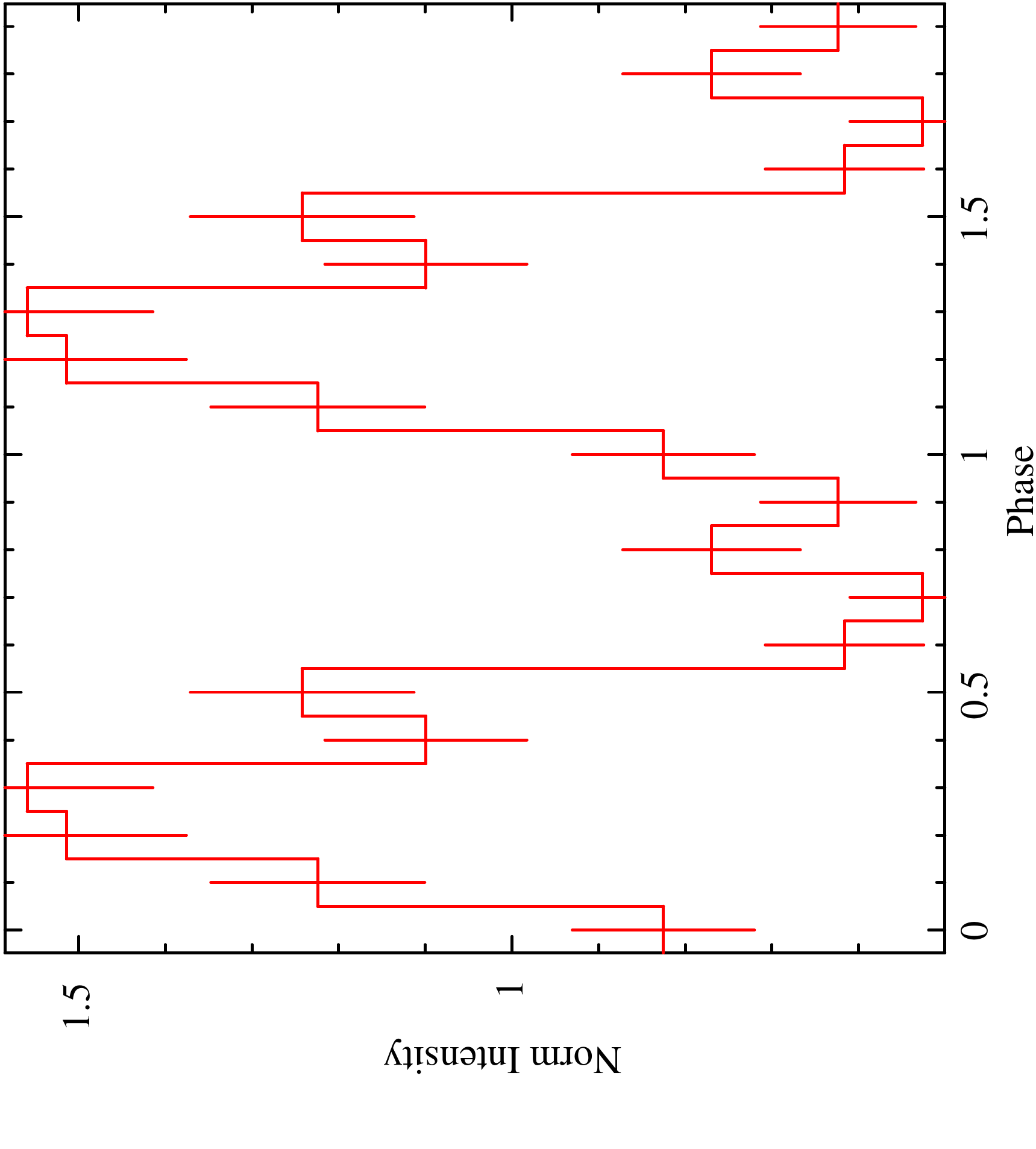}
}
\caption{Left: Combined eROSITA Obs.\,1 pulse profile in the energy range 0.2--10.0\,keV. Centre: Same for eROSITA Obs.\,3. Right: Combined \xmm/EPIC pulse profile from 0.5--10.0\,keV. The light curves were folded with the respective best-fit pulse period inferred from the individual observation (see Sect.\,\ref{subsec:timing}).}
   \label{figcommpf}
\end{figure*}

A combined EPIC light curve binned at 506\,s shows that the source flux remains constant with a decreasing trend towards the end of the observation (Fig.~\ref{figlcxmm}). The source displayed an average count rate of 0.018$\pm0.001$ \cts in the energy range of 0.2--12.0\,keV which decreased to 0.001$\pm0.002$ \cts towards the end of the observation.
A Lomb-Scargle periodogram analysis from the combined light curve from the three EPIC instruments confirms a spin period of $\sim$2024\,s, albeit at a much lower flux level than that of the previous \xmm and the eROSITA observations studied in this work. The pulse  profile is single peaked and the pulsed fraction integrated over the entire \xmm band is $\sim$40\% (Fig.~\ref{figcommpf}), similar to the estimated values using eROSITA data. 


\subsection{X-ray spectral analysis}
Only eROSITA  Obs.\,1, 3, and 4 were used to perform spectral analysis due to insufficient statistics of Obs.\,2 as mentioned before.
For the spectral analysis of the \xmm observation, the SAS tasks \texttt{rmfgen} and \texttt{arfgen} were used to create the redistribution matrix and ancillary file and the eSASS task \texttt{srctool} was used for the same purpose in the case of the eROSITA observations.
eROSITA spectra were binned to achieve a minimum of 20 counts per spectral bin using the $\chi^{2}$-statistic. The \xmm/EPIC spectrum was binned to achieve a minimum of one count per spectral bin and C-statistic was used because of the lower statistics. Errors were estimated at 90\% confidence intervals. The spectral analysis was performed using the {\small XSPEC} fitting package, version~12.10.1 \citep{1996ASPC..101...17A}.
\label{subsec:specr}

\subsubsection{Obs.\,1}
The X-ray spectrum during Obs.\,1 was investigated by performing a simultaneous fit using data from TM3 and TM4 with an inter-calibration constant set free. A simple absorbed power-law model provided an adequate description of the spectrum. The X-ray absorption was modelled using the {\texttt tbabs} model \citep{2000ApJ...542..914W} with atomic cross sections adopted from \cite{1996ApJ...465..487V}.
For this purpose we used two absorption components: one to describe the Galactic foreground absorption and another to account for the column density of both the interstellar medium of the LMC and the intrinsic absorption corresponding to \src.
For this absorption component, the abundances were set to 0.5 solar for elements heavier than helium. For the Galactic photo-electric absorption, we used a fixed column density \citep{1990ARA&A..28..215D} with abundances taken from \cite{2000ApJ...542..914W}.

\begin{figure*}
\centerline{
  \hspace{-0.6cm}
  \includegraphics[width=0.37\textwidth,angle=-90]{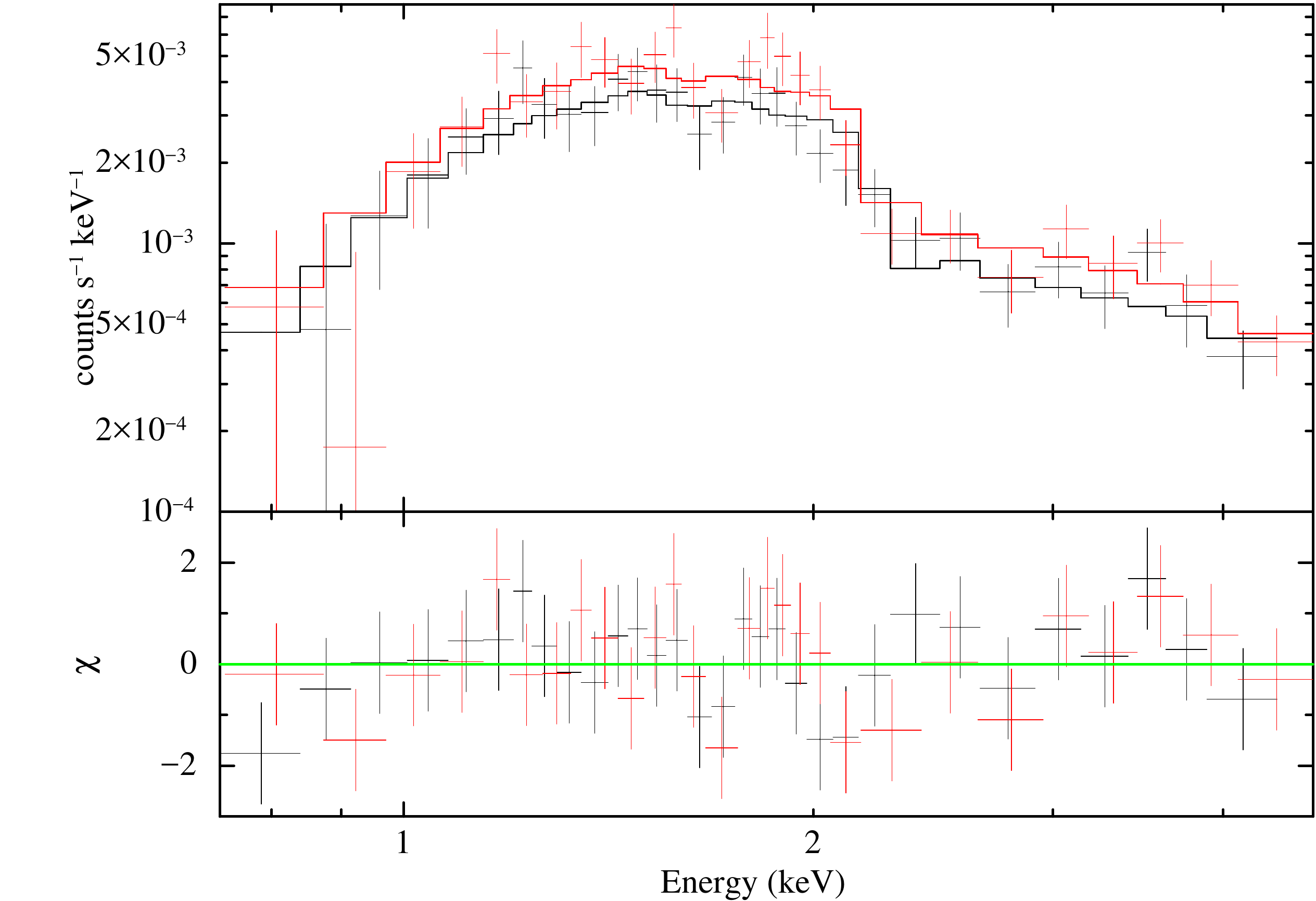}
  \hspace{-0.6cm}
  \includegraphics[width=0.37\textwidth,angle=-90]{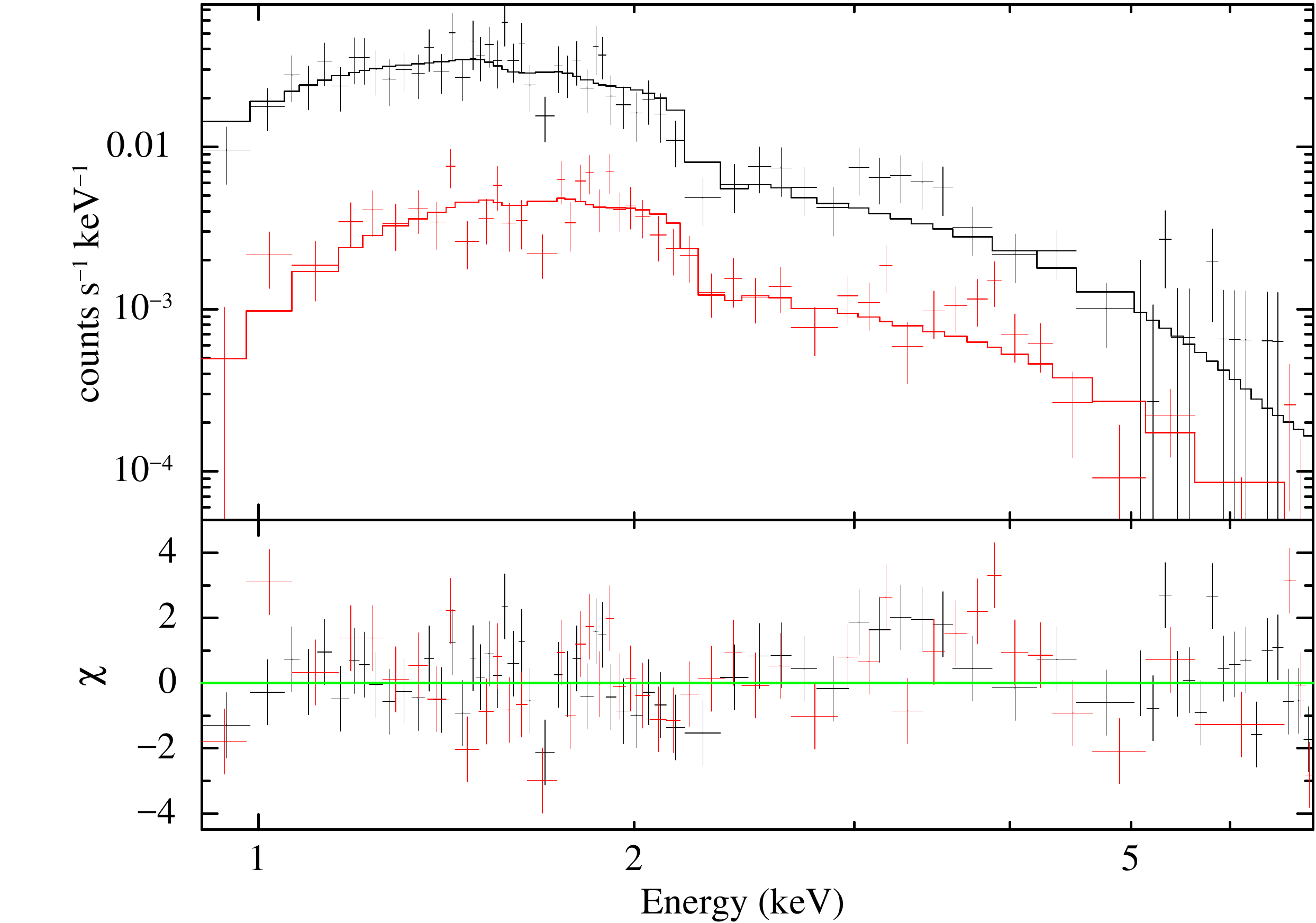}
}
\caption{eROSITA spectra of \src from Obs.\,1. 
         Left: Average spectra are shown from TM3 (black) and TM4 (red) together with the best-fit model as histograms. 
         Right: Simultaneous spectral fit from the flaring (black) and non-flaring (red) intervals using combined spectra from the two telescope modules. 
         For both cases, the residuals are plotted in the lower panels.}
   \label{figspecobs1-int}
\end{figure*}

The spectral parameters for the best-fit model using an absorbed power law are listed in Table~\ref{tab:spectra} and the spectra and best-fit model are shown in Fig.~\ref{figspecobs1-int} (left). The source was detected with an average absorption-corrected luminosity of 7\ergs{35} which indicates a similar intensity to that reported in \cite{2018MNRAS.475..220V}. However, the addition of a black-body component or a partial covering absorber does not significantly improve the spectral fit, as was reported by \citet{2018MNRAS.475..220V}. 
 
The hardness ratio plots in Fig.~\ref{figcommlc} indicate a spectral hardening immediately after the flaring intervals with a dip seen during the flaring intervals. In order to investigate this further, a spectrum was extracted (by combining TM3 and TM4) for the combined flare intervals marked with red dashed lines in Fig.~\ref{figcommlc} and was compared with the corresponding spectrum for the remaining part of the observation. The spectral analysis was performed initially with all the spectral parameters tied, except the power-law normalisations which were set free. The results show a steady increase in power-law normalisation from the `non-flaring' to the `flaring' interval indicating a luminosity $>$ 1\ergs{36} during the flares and the an average luminosity of $\sim$2\ergs{35} during the rest of the intervals. 
Additionally, the fit statistics improved significantly by setting the N$_{\rm H}$ free, which was found to be smaller during the flaring interval by a factor of approximately two. 
The spectra at different intensity levels are shown in Fig.~\ref{figspecobs1-int} (right) and the best-fit parameters of the simultaneous fit are given in Table~\ref{tab:intspectra}. 
The observed scenario is consistent with that seen in typical Galactic SFXTs, where the N$_{\rm H}$ increases just after a flare following a drop during the flare peak. The picture is in agreement with the idea that the flares/outbursts are triggered
by the presence of dense structures in the wind interacting with the X-rays from the compact object \citep[leading to photoionisation]{refId0}.

\subsubsection{Obs.\,3}

\begin{figure*}
\centerline{
  \hspace{-0.6cm}
  \includegraphics[width=0.37\textwidth,angle=-90]{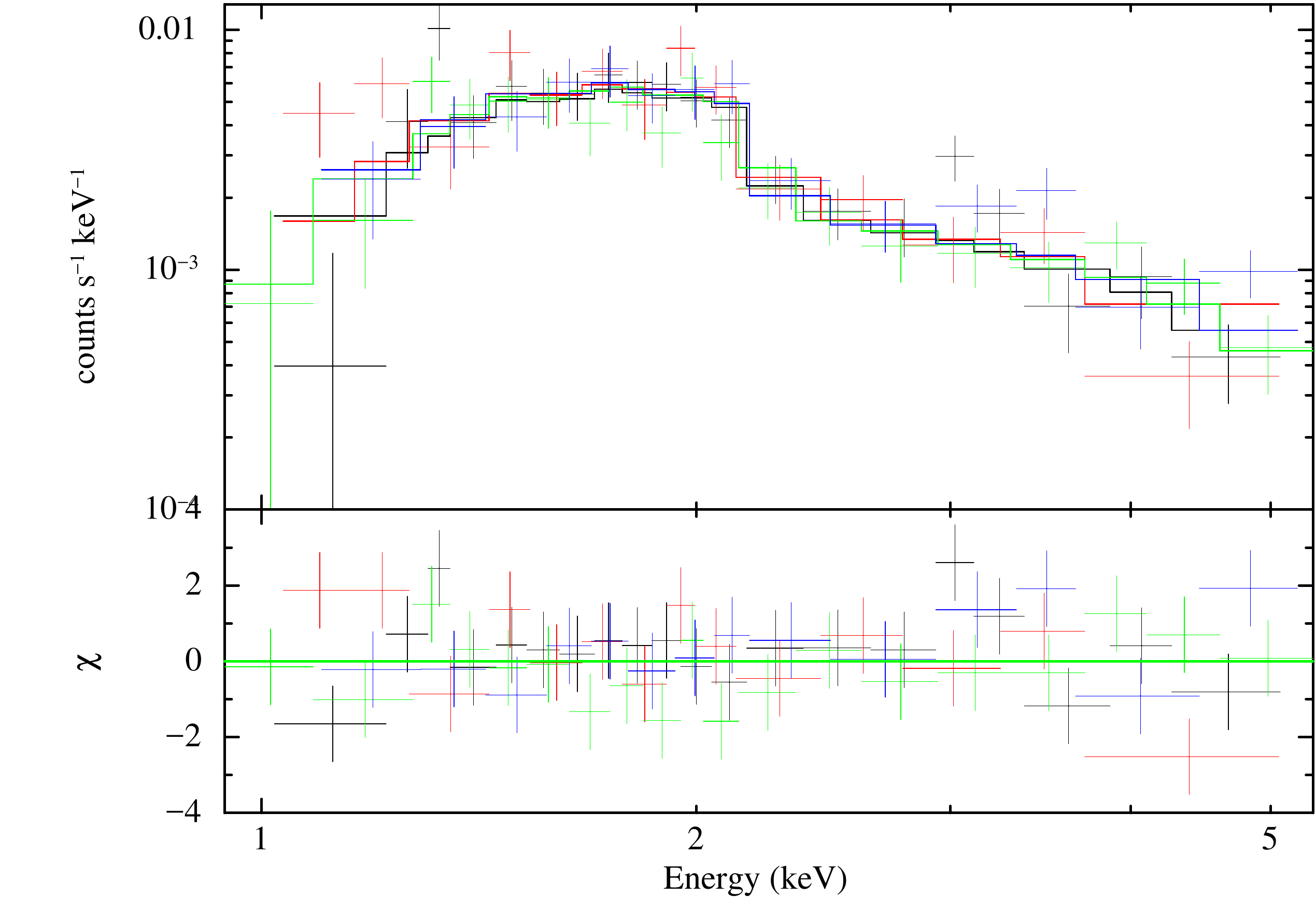}
  \hspace{-0.6cm}
  \includegraphics[width=0.37\textwidth,angle=-90]{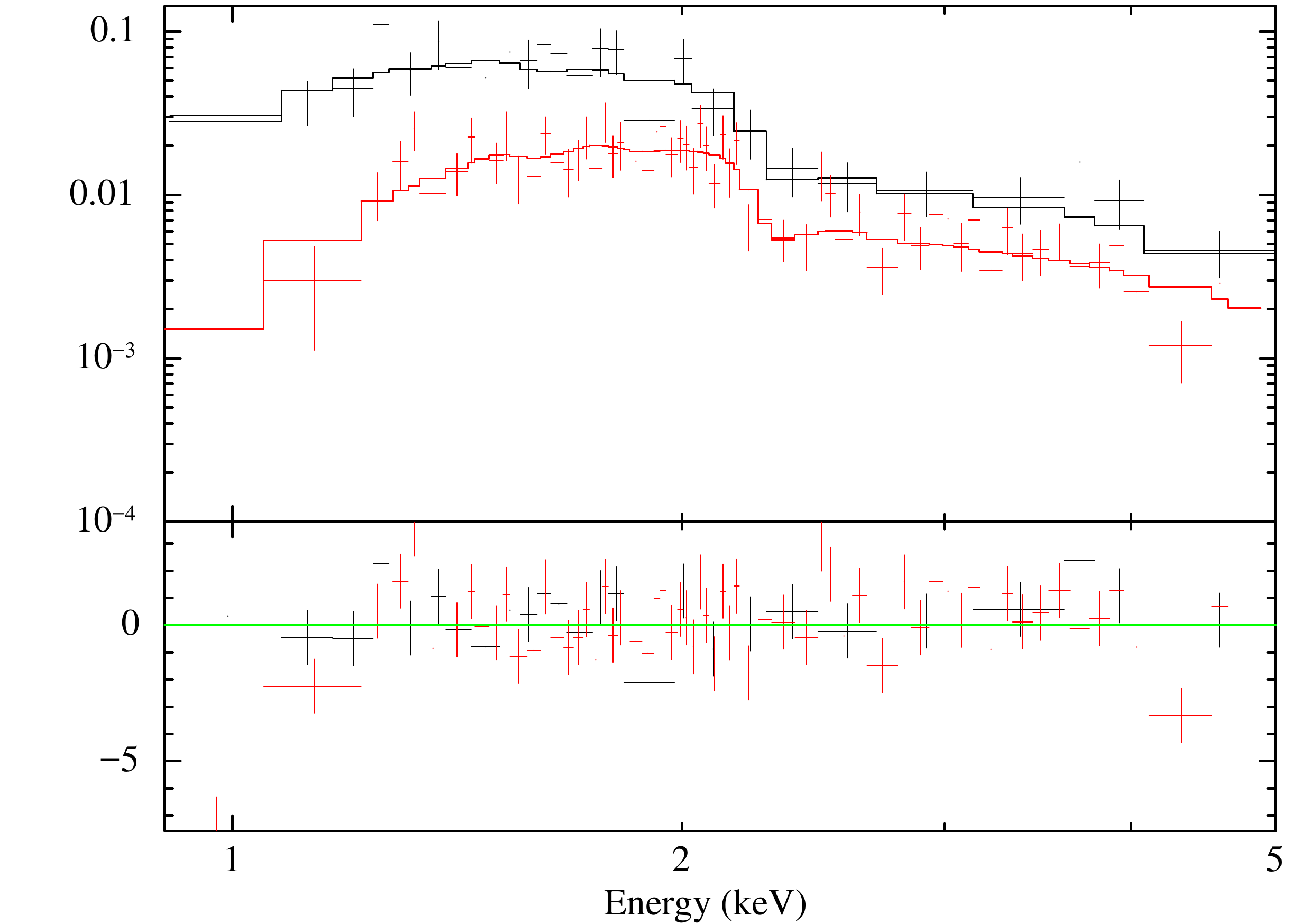}
}
\caption{eROSITA spectra of \src from Obs.\,3. 
         Left: Average spectra are shown from TM1 (black), TM2 (red), TM3 (green), and TM6 (blue) together with the best-fit model as histograms. 
         Right: Simultaneous spectral fit from the flaring (black) and non-flaring (red) intervals using combined spectra from the four telescope modules. 
         For both cases, the residuals are plotted in the lower panels.}
   \label{figspecobs2-int}
\end{figure*}

Spectral analysis of Obs.\,3 was performed using the same spectral model as above and resulted in similar spectral parameters as inferred from  Obs.\,1. \src however remained at a higher average luminosity and exhibited a marginally higher local absorption. An intensity-resolved spectral analysis was performed (spectra from all TMs except TM4 were combined, and TM4 was not used) to compare the spectrum of the flare seen at the end of the observation with the non-flare spectrum. The flare interval exhibited a higher power-law normalisation and significantly lower N$_{\rm H}$ consistent with the behaviour during Obs.\,1 and SFXTs in general during flares. The source reached a peak luminosity of $\sim$ 2\ergs{36} during the flaring interval and displayed an average luminosity of $\sim$ 8\ergs{35} during the rest of the observation.
The spectra at different intensity levels are shown in Fig.~\ref{figspecobs2-int} (right) and the best-fit parameters of the simultaneous fit are listed in Table~\ref{tab:intspectra}.

\subsubsection{\xmm}
\label{Sect:xmm}
An absorbed power-law model provided a good fit to the data with the different absorption components as described in the analysis of the eROSITA spectra. \xmm observed the source in January 2018 in a much fainter state with an observed luminosity of L$_{x}$ of 2\ergs{34} (0.2--10.0\,keV). The decrease in count rate towards the end of the observation (see Sect.~\ref{subsec:timing}) indicates L$_{\rm x}$ $\approx$\oergs{33}.

\begin{figure}
\centerline{
  \hspace{-0.5cm}
  \includegraphics[width=0.7\columnwidth,angle=-90]{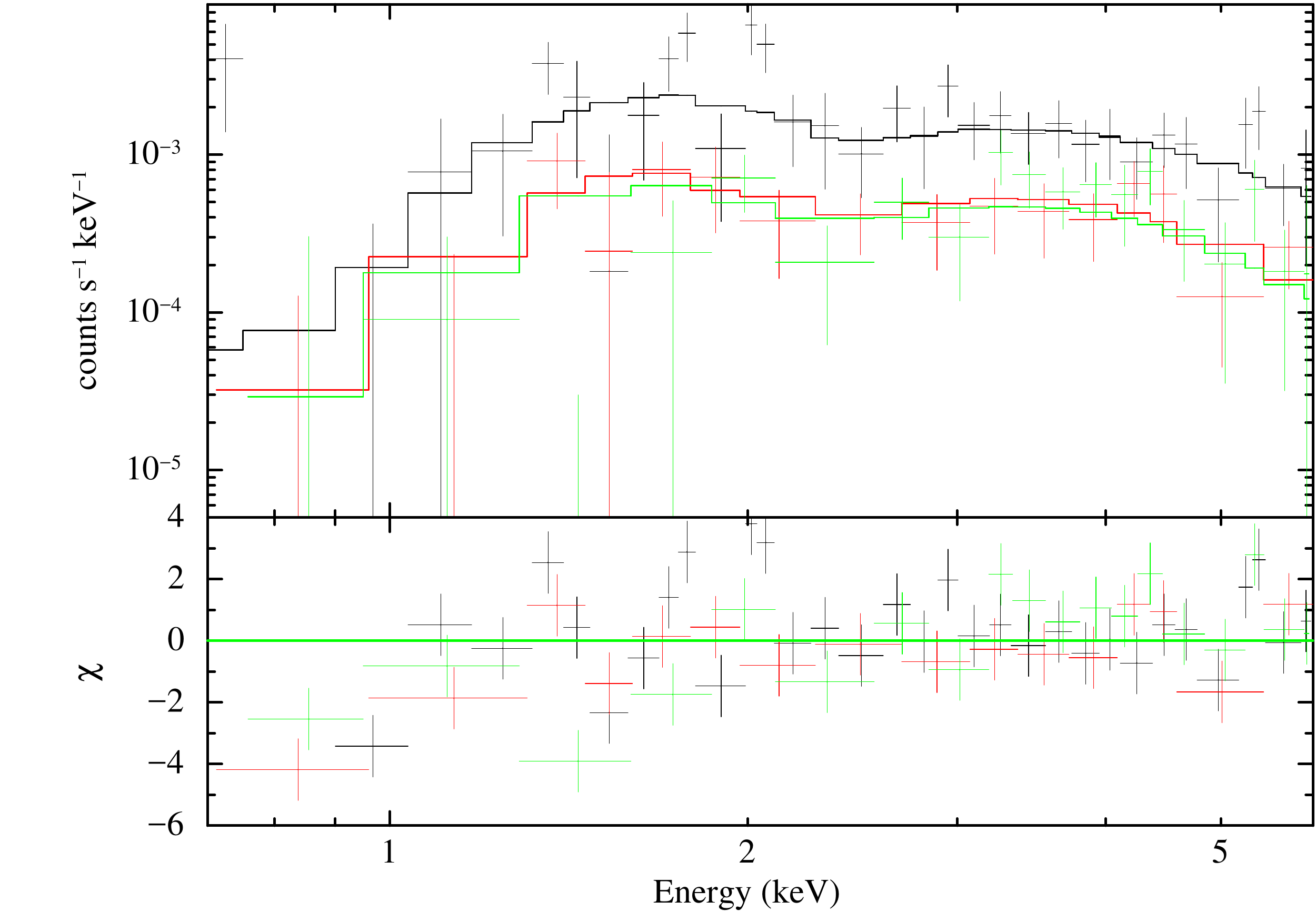}
}
\caption{\xmm/EPIC spectra of \src from observation 0822310101 with the best-fit model. The pn, M1 and M2 are shown in black, red and green, respectively.}
   \label{figspecxmm}
\end{figure}

Figure~\ref{figspecxmm} shows the EPIC-pn, MOS1, and MOS2 spectra with best-fit model derived from a simultaneous analysis. The best-fit parameters are summarised in Table~\ref{tab:spectra}. 

\begin{table}
\caption{X-ray spectral modelling of observation-averaged X-ray spectra.}
\scalebox{0.9}{
\begin{threeparttable}
\begin{tabular}{lccc}
\multicolumn{4}{l}{\xmm constant*TBabs*TBvarabs*powerlaw}\\
\hline
Component & Parameter & Value & units \\
\hline\noalign{\smallskip}
TBabs\tablefootmark{a}    &  N$_{\rm H}^{(1)}$ &  0.0618 (fixed)         &  \ohcm{22}    \\ \noalign{\smallskip}  
TBvarabs                  &  N$_{\rm H}^{(2)}$ &  3.1$^{+1.8}_{-1.3}$    &  \ohcm{22}    \\ \noalign{\smallskip}  
powerlaw                  &  $\Gamma$    &  1.1$\pm0.5$            &     -         \\ \noalign{\smallskip}                 
\hline \noalign{\smallskip}   
\multicolumn{2}{l}{Observed Flux\tablefootmark{b}}   &    1.0$^{+0.2}_{-0.6}$    &  \oergcm{-13}    \\ \noalign{\smallskip}   
\multicolumn{2}{l}{Luminosity\tablefootmark{c}}      &    2.8 (3.7)              &  \oergs{34}      \\ \noalign{\smallskip}    
\hline \noalign{\smallskip}
                          & $cstat^2_{\rm red}$/DOF  &     411/467               &                  \\ \noalign{\smallskip} 
\hline \noalign{\smallskip}
\hline\noalign{\smallskip}
\multicolumn{4}{l}{eROSITA Obs.\,1 constant*TBabs*TBvarabs*powerlaw}\\
\hline
Component & Parameter & Value & units \\
\hline\noalign{\smallskip}
TBabs\tablefootmark{a}    &   N$_{\rm H}^{(1)}$      &  0.0618 (fixed)               &  \ohcm{22}    \\ \noalign{\smallskip}  
TBvarabs                  &    N$_{\rm H}^{(2)}$    &  1.2$\pm$0.4                  &  \ohcm{22}    \\ \noalign{\smallskip}  
powerlaw                  &  $\Gamma$ &  0.33$^{+0.29}_{-0.27}$                &     -         \\ \noalign{\smallskip}                 
\hline \noalign{\smallskip}   
\multicolumn{2}{l}{Observed Flux\tablefootmark{b}}      &     2.3$^{+0.4}_{-0.4}$     &  \oergcm{-12}   \\ \noalign{\smallskip}   
\multicolumn{2}{l}{Luminosity\tablefootmark{c}}         &    7.1                &  \oergs{35}     \\ \noalign{\smallskip}    
\hline \noalign{\smallskip}
         & $\chi^2_{\rm red}$/DOF &     0.84/55             &            \\ \noalign{\smallskip} 
\hline \noalign{\smallskip}
\hline\noalign{\smallskip}
\multicolumn{4}{l}{eROSITA Obs.\,3 constant*TBabs*TBvarabs*powerlaw}\\
\hline
Component & Parameter & Value & units \\
\hline\noalign{\smallskip}
TBabs\tablefootmark{a}    &  N$_{\rm H}^{(1)}$ &  0.0618 (fixed)            &  \ohcm{22}    \\ \noalign{\smallskip}  
TBvarabs                  &  N$_{\rm H}^{(2)}$ &  3.1$^{+0.8}_{-0.7}$       &  \ohcm{22}    \\ \noalign{\smallskip}  
powerlaw                  &  $\Gamma$    &  0.93$^{+0.34}_{-0.32}$    &     -         \\ \noalign{\smallskip}                 
\hline \noalign{\smallskip}   
\multicolumn{2}{l}{Observed Flux\tablefootmark{b}}       &     2.6$^{+0.5}_{-0.5}$   &  \oergcm{-12} \\ \noalign{\smallskip}   
\multicolumn{2}{l}{Luminosity\tablefootmark{c}}          &     1                     &   \oergs{36}  \\ \noalign{\smallskip}
\hline \noalign{\smallskip}
                                         & $\chi^2_{\rm red}$/DOF &     1.03/67      &               \\ \noalign{\smallskip} 
\hline \noalign{\smallskip}
\end{tabular}
\tablefoot{
\tablefoottext{a}{Contribution of Galactic foreground absorption, column density fixed to the weighted average value of \citet{1990ARA&A..28..215D}.}
\tablefoottext{b}{Observed flux for the best-fit model in the 0.2-10.0\,keV energy band. In the case of eROSITA Obs.\,1, the normalisation constant for TM4 with TM3 fixed to 1 is 1.28$^{+0.13}_{-0.15}$. In the case of eROSITA Obs.\,3 the values are 1.00$^{+0.15}_{-0.17}$, 0.87$^{+0.12}_{-0.13}$, 1.01$^{+0.13}_{-0.12}$ for TM2, TM3 and TM6, respectively (the constant for TM1 fixed to 1). Fluxes reported are the average values inferred from the simultaneous fit.}
\tablefoottext{c}{Absorption-corrected X-ray luminosity (0.2-10.0\,keV), assuming a distance of 50\,kpc.}
}
\end{threeparttable}
}
\label{tab:spectra}
\end{table}

\begin{table*}
\caption{X-ray spectral modelling of intensity-resolved eROSITA spectra.}\begin{center}
\begin{tabular}{lllccl}
\hline\noalign{\smallskip}
Obs. & Component    & Parameter    & Value (Flare)    & Value (Flare removed)    & units \\
\noalign{\smallskip}\hline\noalign{\smallskip}
1 & TBabs        &  N$_{\rm H}^{(1)}$ &    \multicolumn{2}{c}{0.0618 (fixed)}        &  \ohcm{22}      \\  \noalign{\smallskip}
  & TBvarabs     &  N$_{\rm H}^{(2)}$ &  1.3$\pm0.4$      &  2.7$\pm0.6$    &  \ohcm{22}      \\   \noalign{\smallskip}
  & Powerlaw    &  $\Gamma$        &  \multicolumn{2}{c}{0.98$^{+0.28}_{-0.27}$}    &             \\  \noalign{\smallskip}                
  & Constant factor     &        &  1.0 (fixed)              & 0.24$\pm0.04$           &              \\                 
  &              &  F$_{\rm x}$ &  4.1$\pm1.0$              &  0.9$\pm0.2$             &  \oergcm{-12}  \\
  &              &  L$_{\rm x}$ &  1.2              &  0.2             &  \oergs{36}    \\    
\noalign{\smallskip}\hline\noalign{\smallskip}   
3 & TBabs        &  N$_{\rm H}^{(1)}$ &  \multicolumn{2}{c}{0.0618 (fixed)}       &  \ohcm{22}      \\ \noalign{\smallskip}
  & TBvarabs     &  N$_{\rm H}^{(2)}$ &    1.4$\pm0.5$      &  3.5$\pm0.7$   &  \ohcm{22}      \\ \noalign{\smallskip} 
  &  Powerlaw    &  $\Gamma$        &  \multicolumn{2}{c}{0.92$^{+0.28}_{-0.26}$}    &             \\ \noalign{\smallskip}                
  & Constant factor     &        &  1.0 (fixed)              & 0.56$\pm0.10$           &              \\                
  &              &  F$_{\rm x}$ &  4.4$\pm1.0$              &  2.3$\pm0.5$              &  \oergcm{-12}  \\
  &              &  L$_{\rm x}$ &  1.4              &  0.8              &  \oergs{36}    \\    
\noalign{\smallskip}\hline\noalign{\smallskip}
\end{tabular}
\end{center}
\tablefoot{
Results from a simultaneous fit of an absorbed power-law model (TBabs*TBvarabs*powerlaw) to the eROSITA intensity-resolved spectra.\\
Fluxes and luminosities are integrated over the 0.2-10.0 keV energy band. L$_{\rm x}$ are corrected for absorption assuming a distance of 50 kpc.\\
}
\label{tab:intspectra}
\end{table*}



\section{Discussion and Conclusion}
\label{sec:discussion}
We present here a detailed timing and spectral study of the HMXB \src in the LMC using eROSITA and \xmm observations. Pulsations were detected with high confidence at $\sim$2020\,s, confirming its nature as a neutron star. In the literature the optical counterpart of \src\ has been classified either as a  B0 IIIe \citep{2018MNRAS.475.3253V} or B0 IIe-Ib star \citep{2018MNRAS.475..220V} . However, as  pointed out by \citet{2018MNRAS.475.3253V}, this discrepancy could be due to the lower S/N of the blue spectrum. Moreover, \cite{2018MNRAS.475..220V} used a hybrid model based on spectrophotometric criteria. Based on those, the absolute magnitude of the system was at least 1 magnitude brighter than a main sequence star. This indicates that \src\ is a supergiant HMXB in the LMC.

Further, we find strong evidence that the \src\ belongs to the  rare and elusive class of SFXTs that remain in a subluminous state most of the time and exhibit short flaring/outburst activities on timescales of several thousand seconds. This is the first such conclusive evidence outside our Galaxy where the source displays all the characteristics analogous to SFXTs, given its flaring activity as well as a high dynamic luminosity range as expected for SFXTs. The only other indirect evidence obtained before was in the case of IC10 X--2, where the classification was solely based on its high dynamic luminosity range \citep{2014ApJ...789...64L}, and XMMU\,J053320.8-684122, where the classification was based on fast flaring behaviour alone \citep[][]{2018MNRAS.475..220V}.

\subsection{Nature of \src: The first conclusive evidence of an SFXT outside our Galaxy}
The light curves obtained from the eROSITA observations reveal flares lasting several thousand seconds. The source reached a luminosity of >\erg{36} during the peak of the flares which lasted for 9\,ks and 5\,ks during eROSITA Obs.\,1 and 3, respectively. The source remained in an `intermediate' state during the rest of the observation in both cases, with $L_{\rm x}$ at \erg{35}. The average luminosities measured during Obs.\,1 and 3 are also similar to that reported in \cite{2018MNRAS.475..220V}. During eROSITA Obs.\,4, the source displayed an average luminosity measured to be $\sim$20 times lower than during Obs.\,1 and 3 with signatures of flaring activity where $L_{\rm x}$ rose to 4\ergs{36}.
Further, an additional \xmm observation in 2018 indicated that the source remained in a subluminous state of $L_{\rm x}$ $\sim$\erg{34} throughout, with the flux dropping to \erg{33} towards the end, the lowest  reported so far from \src. 

The flares during the eROSITA Obs.\,1 and 3 indicate a dynamic range of $\sim$15--30 with respect to the average luminosity attained during the observations. However, the total observed dynamic luminosity range is  greater than 1000 when considering the ratio between the peak of the bright flare seen during Obs.\,4 of 2$\times$\oergs{36} and the drop in luminosity to  \erg{33} towards the end of the \xmm  observation. 
This is a typical dynamic range observed for SFXTs, with the luminosity during the \xmm observation also being typical for SFXTs during quiescence \cite[see e.g.][]{2010A&A...519A...6B}.

Comparison of the light curves in the soft and hard X-ray bands reveals variations in hardness-ratio which reached its minimum during the peaks of the flares (as seen in eROSITA Obs.\,1 and 3) and increased steeply shortly after the flares (Obs.\,1). This is also supported by a comparison of spectra from the peak `flare' and `non-flare' intervals with a corresponding decrease in N$_{\rm H}$ during the peak of the flare  by a factor of approximately two. This behaviour can be attributed to the presence of massive and dense structures in the winds of the supergiant companions of SFXTs. Such clumps intercepted by the NS can lead to a temporary higher mass accretion rate, giving rise to X-ray flares or outbursts \citep{2011A&A...531A.130B,refId0}. Although gating mechanisms introduced by the spinning NS and its strong magnetic field possibly also play a role in contributing to the total dynamic luminosity range of SFXTs from quiescence to the outbursts/flares \citep[for e.g.][]{Bozzo_2008,Grebenev2007149}. The flares and the corresponding changes in the local absorption are mainly triggered by the presence of clumps approaching the NS. The drop in N$_{\rm H}$
near the flare peak can most likely be ascribed
to the photoionisation of the clump medium. Further, a re-increase in
N$_{\rm H}$ is expected after the flare, which is due to the recombination that is allowed by the decreasing X-ray flux \citep{refId0}.  

\subsection{Physical properties of the clump ingested during flare}
\label{sec:discussion-flare}
 `Clumpy' structures in supergiant winds are believed to be generated by magneto-hydrodynamic instabilities in the winds of supergiants and have been supported by both numerical simulations \citep[see for e.g.][]{2007A&A...476.1331O} and observations of isolated O-B stars \citep[see for e.g.][]{2018A&A...614A..60R} and recently by \cite{2020arXiv200616216E}. The changes in density/velocity in the stellar wind cause a proportional change in X-ray luminosity which in turn affects the stellar wind through ionisation \citep{2015A&A...579A.111K}. In the `clumpy' wind scenario, flares are produced by interception of a clump by the NS which leads to a temporary increase in mass accretion rate. In this case the increase in the local absorption column is also proportional to the size of the accreted structure \citep{2005A&A...441L...1I,2010ASPC..422...57N,2013A&A...556A..30B}. Flares in SFXTs are difficult to catch because of their sporadic nature. Therefore, detailed studies have only been possible for a limited number of flares until today and are restricted to Galactic sources. The observed physical properties of the wind and the properties of the clumps giving rise to the flares are largely unknown. The long exposures of \src available from eROSITA observations provide a unique opportunity to probe the properties of the clumps in great detail. We calculated the physical properties of the possible clump (assuming a spherical geometry) giving rise to the stronger flare seen in Obs.\,1. The flare lasted for 9\,ks and the X-ray luminosity reached a value $L_{\rm x}$ of 1.2\ergs{36} during the peak. The characteristic time for the accretion of the clump ($ t_{\rm flare}$) is assumed to be equal to the duration of the flare. The relative velocity between the clump and the NS is $v \sim 1000$\,km s$^{-1}$ (spherical clump moving with the same velocity as the surrounding wind material). This is a reasonable assumption given that for O-B supergiants the terminal wind velocity is assumed to be in the range of 500-2000\,km s$^{-1}$ and the wind velocity profile can be obtained by assuming the $\beta$-law \citep{2008A&ARv..16..209P}.
 
The radial extent of the clump can be estimated as 
\begin{equation}
R_{\rm c} \simeq v t_{\rm flare}/2 = 5\times10^{11}~{\rm cm}.  
\end{equation}
The  N$_{\rm H}$ measured at the flare peak is $1.3\times$\ohcm{22},
which can be used to estimate the mass density of the clump as 
\begin{equation}
n_{\rm c}\simeq \frac{N_{\rm H}}{R_{\rm c}} = 4\times10^{-14} {\rm g~cm^{-3}}.
\end{equation}
This implies a characteristic volume of the clump of 
$V_{c}\simeq {R_{\rm c}}^{3}\approx 10^{35} \ \rm cm^{3}$, 
which would then give  the clump responsible for the 
bright flare a  mass of $M_{\rm c}\sim 10^{21}$~g. The estimated values of the mass and radius of the clump are compatible with estimates of the clump mass and radius obtained for other supergiant HMXBs \citep{2011A&A...531A.130B,2014A&A...563A..70M,2014MNRAS.442.2691P}. The derived value is also in qualitative agreement with the predictions of the clumpy wind model of HMXBs \citep{2009MNRAS.398.2152D}. However, the estimate of the mass of the clump is approximately 1000 times more massive than predicted from radiative hydrodynamic simulations of hot, massive stars \citep{2018A&A...611A..17S}. This discrepancy has been discussed extensively by \cite{2017SSRv..212...59M}, who point to either current limitations in the hydrodynamic simulations or additional factors (e.g. centrifugal and/or magnetic barriers) playing a role in the release of the X-ray luminosity, thus affecting the estimates of the clump masses.
In the quasi-spherical subsonic accretion regime, the clump material is not directly deposited onto the NS. Instead, matter settles down to form a quasi-static shell above the magnetosphere. The shell mediates the angular momentum transfer with respect to the magnetosphere via viscous stresses and the actual mass accreted onto the NS surface could be significantly smaller and of the order of 10\% of that mass \citep[see][]{10.1111/j.1365-2966.2011.20026.x,10.1093/mnras/stu1027}. The mass accreted per episode could be of the order of $M_{\rm acc}\sim 10^{20}$~g. The luminosity of such accreted episodes (i.e. flares) could subsequently be estimated by considering the characteristic time for the accretion of the clump ($ t_{\rm flare}$). 
The resulting predicted mass accretion rate is
$\dot{M}_{\rm acc}\sim 10^{16}$ g s$^{-1}$, and the luminosity can be estimated by converting the dynamic energy to luminosity, that is,
\begin{equation}
L_{X}\approx G \frac{M_{\rm NS}\dot{M}_{\rm acc}}{R_{\rm NS}}\approx 2\times 10^{36} \rm erg\,s^{-1} 
,\end{equation}
%

which is compatible with that measured from the spectral analysis. It is worth noting that these estimates are prone to large uncertainties due to projection effects related to the viewing angle between the observers line of sight and the direction from which the clump approaches the NS \citep[see for e.g.][]{2010MNRAS.408.1540D}.

\subsection{A highly magnetised slowly rotating neutron star?}
The total dynamic luminosity range of SFXTs, especially the quiescent state luminosity, can be boosted by transitions across centrifugal and magnetic barriers as proposed in \cite{Bozzo_2008}. Operation of the `gating' mechanism requires a slowly spinning (several hundred to several thousand seconds) and highly magnetised neutron star. However, this scenario is difficult to validate as robust spin period and/or magnetic field estimates only exist for a very small number of SFXTs; those with well-established spin periods being IGR J16418--4532 \citep[1212\,s][]{2012MNRAS.420..554S} and IGR J11215--5952 \citep[187\,s][]{2017ApJ...838..133S}. \src,\ with its precisely measured spin period, is a well-established slowly rotating neutron star and provides a unique opportunity to probe the above scenario. Considering that the average luminosity measured from the source is $\sim$\oergs{36} and assuming the source is near spin equilibrium condition at P = 2020\,s, one can estimate the magnetic field B of the neutron star from \cite{2014MNRAS.437.3664H} as $\sim$5$\times10^{14}$\,G. We note that the `fastness' parameter $\omega_{s}$ is assumed to take a value of 1 here. In the scenario of \cite{1979ApJ...234..296G}, with $\omega_{s}=0.35,$ a magnetic field strength of $\sim10^{14}$\,G is predicted.  This indicates a highly magnetised neutron star with a field strength comparable to that of a magnetar. However, it must be noted that these estimates are based on the assumption of the transfer of angular momentum between the NS and the infalling matter being mediated by the accretion disc which is probably not true in the case of SFXTs. Another way to estimate the magnetic field in these systems is to consider an expression for the equilibrium period for quasi-spherical settling accretion which takes into account the binary orbital period, wind velocity near the NS orbital location, and the mass accreted \citep{2012int..workE..22P}. For a typical wind velocity of 1000\,km s$^{-1}$, the estimated $\dot{M}_{\rm acc}$ of 10$^{16}$ g s$^{-1}$, and an orbital period range of 1--10 d expected in SFXTs, the estimated magnetic field of the neutron star is $\sim10^{12}-10^{13}$\,G which points towards typical values estimated for accreting pulsars. 
The cases of wind versus disc accretion with respect to the origin of long-period X-ray pulsars has been discussed extensively in the literature \citep[e.g.][]{2007MNRAS.375..698I}. 
By following Eqs. 7 and 8 of \citet{2007MNRAS.375..698I}, which assume a system in equilibrium, and the same binary orbital period and wind velocity as above, we estimate a magnetic field of $\sim3\times10^{15}$\,G and $\sim2\times10^{13}$\,G for disc and wind accretion, respectively. This demonstrates that regardless of the detailed model assumptions, reaching spin equilibrium for the wind accretion case requires B to be a factor of 100-1000 smaller than for disc accretion.

Finally, although our estimates of the spin period were inconclusive in terms of the spin evolution of the pulsar, future monitoring programs are essential for determining whether the pulsar is spinning up or down. It has been proposed that long-period pulsars with magnetar-like field strengths could exhibit a spin up trend due to the decay of the B field of the NS \citep{2020MNRAS.492..762W}. A future measurement of the spin evolution will therefore provide crucial information about the nature of the NS.


\label{sec:conclusions}

\bibliographystyle{aa} 
\bibliography{ref} 

\begin{acknowledgements}
We thank the referee for useful comments and suggestions.
This work is based on data from eROSITA, the primary instrument aboard SRG, a joint Russian-German science mission supported by the Russian Space Agency (Roskosmos), in the interests of the Russian Academy of Sciences represented by its Space Research Institute (IKI), and the Deutsches Zentrum f{\"u}r Luft- und Raumfahrt (DLR). The SRG spacecraft was built by Lavochkin Association (NPOL) and its subcontractors, and is operated by NPOL with support from the Max Planck Institute for Extraterrestrial Physics (MPE).
The development and construction of the eROSITA X-ray instrument was led by MPE, with contributions from the Dr. Karl Remeis Observatory Bamberg \& ECAP (FAU Erlangen-N{\"u}rnberg), the University of Hamburg Observatory, the Leibniz Institute for Astrophysics Potsdam (AIP), and the Institute for Astronomy and Astrophysics of the University of T{\"u}bingen, with the support of DLR and the Max Planck Society. The Argelander Institute for Astronomy of the University of Bonn and the Ludwig Maximilians Universit{\"a}t Munich also participated in the science preparation for eROSITA.
The eROSITA data shown here were processed using the eSASS/NRTA software system developed by the German eROSITA consortium.
This work uses observations obtained with \xmm, an ESA science mission with instruments and contributions directly funded by ESA Member States and NASA. The \xmm project is supported by the DLR and the Max Planck Society.
GV is supported by NASA Grant Number  80NSSC20K0803, in response to XMM-Newton AO-18 Guest Observer Program. GV acknowledges support by NASA Grant number 80NSSC20K1107. 
\end{acknowledgements}

\end{document}